# Hardware-Aware Model Design and Training of Silicon-based Analog Neural Networks.

Giulio Filippeschi, Mirko Brazzini, Cristhopher Mosquera, Marco Lanuzza, Senior Member, IEEE, Alessandro Catania, Senior Member, IEEE, Sebastiano Strangio, Senior Member, IEEE, and Giuseppe Iannaccone, Fellow, IEEE

*Abstract*— Silicon-based analog neural networks physically embody the ideal neural network model in an approximate way. We show that by retraining the neural network using a physics-informed hardware-aware model one can fully recover the inference accuracy of the ideal network model even in the presence of significant non-idealities. This is way more promising for scalability and integration density than the default option of improving the fidelity of the analog neural network at the cost of significant energy, area, and design overhead, through extensive calibration and conservative analog design.

We first present a physics-informed hardware-aware model for a time-domain vector–matrix multiplier implemented with single-transistor floating-gate memory cells that explicitly accounts for two dominant non-idealities of the physical implementation—capacitive crosstalk and bit-line voltage drop—and integrates seamlessly with modern deep-learning workflows. The model discretizes each operation into adaptive time slots, processes activation patterns in parallel, and accumulates their contributions to predict effective multiplier outputs. Using measurements from a 16×16 silicon array, we calibrate the model, show that crosstalk is layout-dependent and often dominant, and introduce an improved weight-extraction procedure that doubles signal-to-error ratio versus an ideal vector-matrix multiplier model.

Finally, we show that by training silicon-based analog neural networks using an hardware aware model in the forward pass we can recover the accuracy of the ideal software networks across three architectures —custom MLP on low-resolution MNIST, LeNet-5 on MNIST, and a VGG-style CNN on CIFAR-10— establishing a complete design-to-deployment workflow for time-domain analog neuromorphic chips.

*Index Terms*— hardware-aware training; time-domain analogue computing; floating-gate memories; in-memory computing; analog neural networks.

## I. INTRODUCTION

This work has been submitted to the IEEE for possible publication. Copyright may be transferred without notice, after which this version may no longer be accessible.

Paper submitted on December 8, 2025. This work was partially supported by the EC Horizon 2020 Research and Innovation Programme under GA AUTOCAPSULE 952118, by Quantavis s.r.l., and by the Italian MUR under the Forelab project of the "Dipartimenti di Eccellenza" programme. U.S. Department of Commerce under Grant 123456." *(Corresponding author: Giuseppe Iannaccone giuseppe.iannaccone@unipi.it)*. Giulio Filippeschi and Mirko Brazzini contributed equally to this work. Giulio Filippeschi, Mirko Brazzini, Alessandro Catania, Sebastiano Strangio and Giuseppe Iannaccone are with the University of Pisa, Dipartimento di Ingegneria dell'Informazione, Via Caruso 16, 56126 Pisa, Italy. Cristhopher Mosquera and Marco Lanuzza are with the Department of Computer Engineering, Modeling, Electronics and Systems, University of Calabria, 87036 Rende, Italy.

The increasing computational demands of modern artificial intelligence systems have exposed the fundamental limitations of the von Neumann architecture, where memory and processing units are physically separated. The constant shuttling of data between memory and compute engines is a dominant energy and latency bottleneck, limiting performance and scalability.

Analog In-Memory Computing (AIMC) has emerged as a compelling alternative to conventional digital neural network accelerators by physically embedding computing primitives in the memory array. A vector-matrix multiplier (VMM) is directly implemented in a AIMC array, where the weights are physically instantiated, leveraging device physics and circuit properties such as Kirchhoff's laws. This approach minimizes data movement, maximizes parallelism since all multiply-and-accumulate (MAC) operations are performed simultaneously, and enables relevant energy efficiency gains for inference tasks.

A broad spectrum of device technologies has been explored for AIMC implementations, ranging from resistive memories such as Phase-Change Memory (PCM) [1], [2], Resistive Random Access Memory (RRAM) [3], and Magnetic RAM (MRAM) [4], to CMOS-based charge storage elements [5], [6], [7], [8], [9], [10]. Each technology presents a specific tradeoff among density, analog programmability, retention, variability, and peripheral circuit requirements.

AIMC accelerators – regardless of the memory technology – typically follow the design-to-deployment flow illustrated in Fig. 1a: a neural network is trained in software, its weights are mapped onto the analog memory array with a program-and-verify procedure, and the inference in the hardware implementation is expected to replicate that of the software model. However, analog hardware inevitably departs from this idealized model. Device-level variability, parasitic coupling, interconnect voltage drops, bitline voltage drops, finite precision in digital-to-analog and analog-to-digital converters, and stochastic programming errors introduce discrepancies compared to the idealized software model. Time-dependent effects, such as conductance drift or pulse-history dependence, further exacerbate the deviations from the ideal behavior. As a result, directly deploying software-trained weights onto analog hardware often leads to inference accuracy degradation. One can mitigate non-idealities through extensive calibration and conservative analog design, at the cost of significant energy, area, and design overhead [7].

**Hardware-aware (HWA) model training** provides a far more scalable and resource-efficient approach: after the first training of the ideal software model, the neural network is retrained using a HWA forward propagation model that explicitly includes the effect of hardware non-idealities in silicon, as depicted in Fig. 1b. In this way, HWA model retraining compensates for device and circuit imperfections, and has proven to be effective across several AIMC platforms [11], [12], [13], [14], [15], [16].

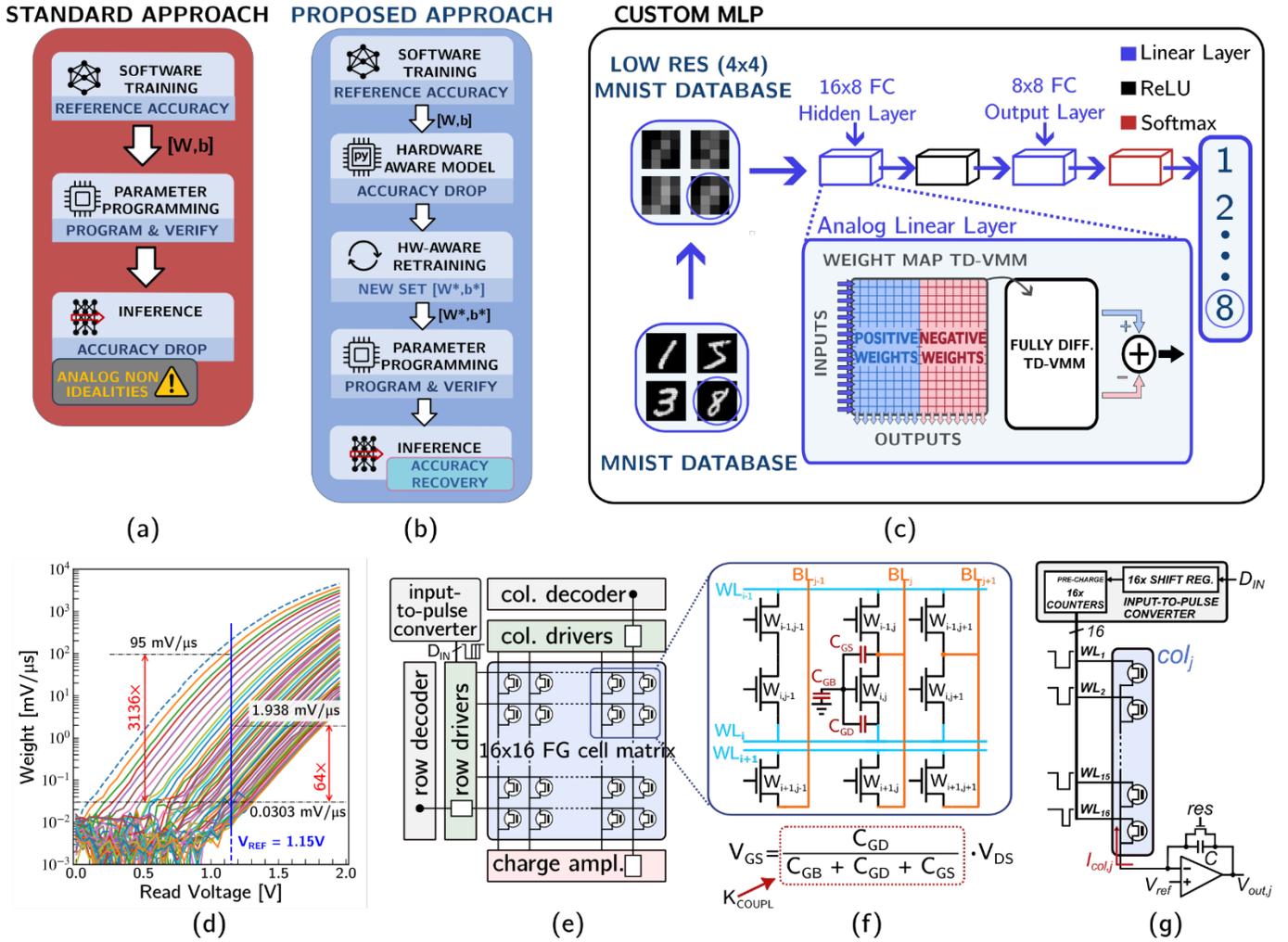

*Figure 1:* Design-to-deployment flow of hardware neural networks (a) Standard approach based on directly transferring on chip the weights trained on a theoretical VMM model. (b) Proposed approach based on retraining the weights with a hardware-aware model before transferring them to the hardware VMM. (c) Multi-layer neural network trained on a low-resolution (4×4) MNIST dataset used for experimental validations. (d) Weight-storing memory cell (1T-FG) characteristics: weight versus read voltage under different program conditions. (e) TD-VMM circuit architecture. (f) 1T-FG cell gate biasing based on capacitive coupling. (g) Details of a TD-VMM single-column.

Existing HWA models share a fundamental assumption: the MAC operation is performed in a single analog step. On the other hand, time-domain (TD) analog computing encodes inputs as variable-width pulses, and the physical MAC is intrinsically executed as time progresses. None of the mentioned HWA frameworks captures time and input-dependent non-idealities occurring during the physical VMM operation. This makes the TD systems highly sensitive to dynamic nonlinearities: the effective weight parameters applied during a single VMM operation depend on the temporal structure of the input vector itself. These input-dependent distortions fundamentally violate the standard abstraction used in current HWA tools and therefore remain unmodeled in the state-of-the-art approaches. Time-dependent distortion modeling is particularly relevant for TD accelerators based on floating-gate (FG) memory cells.

We have recently proposed a TD-VMM [6], exploited to realize a simple multi-layer neural network (concept depicted in Fig. 1c) [7], with weight parameters stored in single transistor floating-gate (1T-FG) memory cells. 1T-FG are implemented with only a two-terminal scheme since the FG biasing ($V_{FG}$) is achieved through capacitive coupling between the applied drain-to-source voltage $V_{DS}$ and $V_{FG}$. Due to their operating principle, FG and flash-based arrays inherently suffer from electrostatic interference, crosstalk, and stringent design constraints affecting charge amplifiers, ADCs, and peripheral circuits all of which limit the performance of the analog front-end during TD operation [17]. In TD-VMMs, these parasitic phenomena directly alter the instantaneous $V_{DS}$ and $V_{FG}$ voltages, generating a nonlinear, input-pattern-dependent distortion whose magnitude depends on the activity level of neighboring lines during each computation window. These distortions affect both the effective programmed weights and neural network inference, since the effective current generated by each FG device during practical use deviates from the one measured during single memory device characterization. In practice, the nominal weights underestimate the effective weights whenever multiple inputs overlap in time, leading to systematic errors that compound as the neural network depth increases.

Here, we propose a HWA model and show that a physics-informed approach including time-dependent nonlinear mechanisms enables a substantial improvement of the analog hardware neural network operation. We obtain the following results:

1) Using our proposed HWA model, we double the precision of experimental evaluation of the effective programmed weights (the "weight extraction") compared to standard procedures based on the idealized VMM behavior.
2) We demonstrate that using the HWA in the forward pass of the training phase of the TD analog neural network, we can reach the **same inference accuracy** of the ideal neural network model. This is a key result of this paper: including the HWA model in the training loop compensates for the non-idealities and the distortions of the analog implementation. In other words, we can accept a non-ideal analog implementation without losing accuracy, if we account for such non-idealities in the training phase. We show that this concept is successfully verified for three neural networks of increasing complexity, tested on MNIST [18] and CIFAR-10 [19] datasets, corroborating findings in amplitude-domain frameworks [16].

This establishes a complete design-to-deployment workflow from physical modeling to weight extraction and network training, tailored to TD analog accelerators and compatible with modern deep learning frameworks. In Section II we discuss the non-idealities of the fabricated TD-VMM circuit. Section III describes the HWA model and present the calibration results. Section IV describes the weight extraction procedure and the HWA model improvement, while Section V demonstrates its application to training strategies and evaluates their effectiveness. Finally, Section VI concludes the paper with a discussion on scalability and perspectives for future neuromorphic hardware.

## II. THE VMM CIRCUIT ARCHITECTURE: OPERATION AND PARASITIC PHENOMENA

The TD-VMM considered in this work, schematically shown in Fig. 1e, is implemented using a 16×16 array of single-poly 1T-FG memory cells together with 16 charge amplifiers, one per column. 1T-FG devices are operated in the subthreshold regime, where the drain current exhibits exponential sensitivity to the FG charge. This operating region enables fine-grained analog programmability [6], low-voltage operation, and compact circuit design, making the devices well suited to serve as a weight-storing tunable current source in an AIMC architecture.

In this 1T-FG cell configuration, the stored weight is proportional to the drain current measured under a constant drain-to-source voltage $V_{ref}$ during the active portion of the input pulse. Weights can be programmed and erased with fine granularity by applying voltage pulses of varying amplitude to the drain terminal while grounding the source, enabling continuous-valued conductance tuning through FG charge injection (Fig. 1d). Once the array is programmed, computation proceeds by stimulating each wordline (WL) with an active-low pulse with time duration proportional to the input element value (Fig. 1e-g). Due to capacitive coupling between the $V_{DS}$ and the $V_{FG}$ (Fig. 1f), each activated device produces a rectangular current waveform whose amplitude is proportional to the programmed weight, while the time integral of this current encodes the physical product between the input pulse width and the stored weight. Within each column, all devices are connected through their drain terminals to a shared bitline (BL) feeding the inverting input of a dedicated charge amplifier (Fig. 1g). The amplifier accumulates the contributions from all active cells on its feedback capacitor $C$, generating the following output:

$$\Delta V_{\text{out},j} = V_{out,j} - V_{reset,j} = \frac{1}{C}\sum_i I_{i,j} \cdot t_i, \quad (1)$$

Where $t_i$ is the $i$-th element of the input (the time duration of the input pulse on the $i$-th wordline), $\Delta V_{\text{out},j}$ is the voltage variation at the output of the charge amplifier of the $j$-th bitline, and $I_{i,j}$ is the current through the memory cell at the crossing point between the $j$-th bitline and the $i$-th wordline when the $i$-th wordline is active (i.e., the "weight"). Signed weights are implemented using a differential mapping [7].

Despite its compact and energy-efficient architecture, the TD-VMM exhibits intrinsic non-idealities. Two effects play a significant role:

1. **Bit Line Voltage Drop:** when the total BL current increases, the finite transconductance of the charge amplifier prevents the BL from maintaining a perfect virtual ground. A parasitic BL voltage drop develops proportionally to the total column current, reducing the effective $V_{DS}$ of the active 1T-FG cells, resulting in a systematic underestimation of the weights under multi-WL activation.

2. **Capacitive Crosstalk Effect:** parasitic capacitive coupling between adjacent WLs and FGs causes the effective gate-source voltage $V_{GS}$ to drop when multiple WLs are activated simultaneously. Because the 1T-FG transistors operate in subthreshold, even modest reductions in $V_{GS}$ lead to significant deviations from nominal weights. Consequently, the total BL current during concurrent WL activity is lower than the sum of the individually measured currents, producing another systematic weight underestimation.

*Both effects are inherently input-dependent*, meaning that the distortion experienced during a VMM operation depends on the number of active input lines, their temporal overlap, and the statistical distribution of pulse widths. As a result, even perfectly programmed devices yield nonlinear and data-dependent MAC results. To characterize these effects experimentally and validate the proposed HWA modeling approach, we have used a fabricated chip integrating two independent 16×16 TD-VMM arrays, exploited to implement the two fully-connected layers of a custom Multi-Layer Perceptron (MLP) trained with a low-resolution (4×4) version of the MNIST dataset, as illustrated in Fig. 1c.

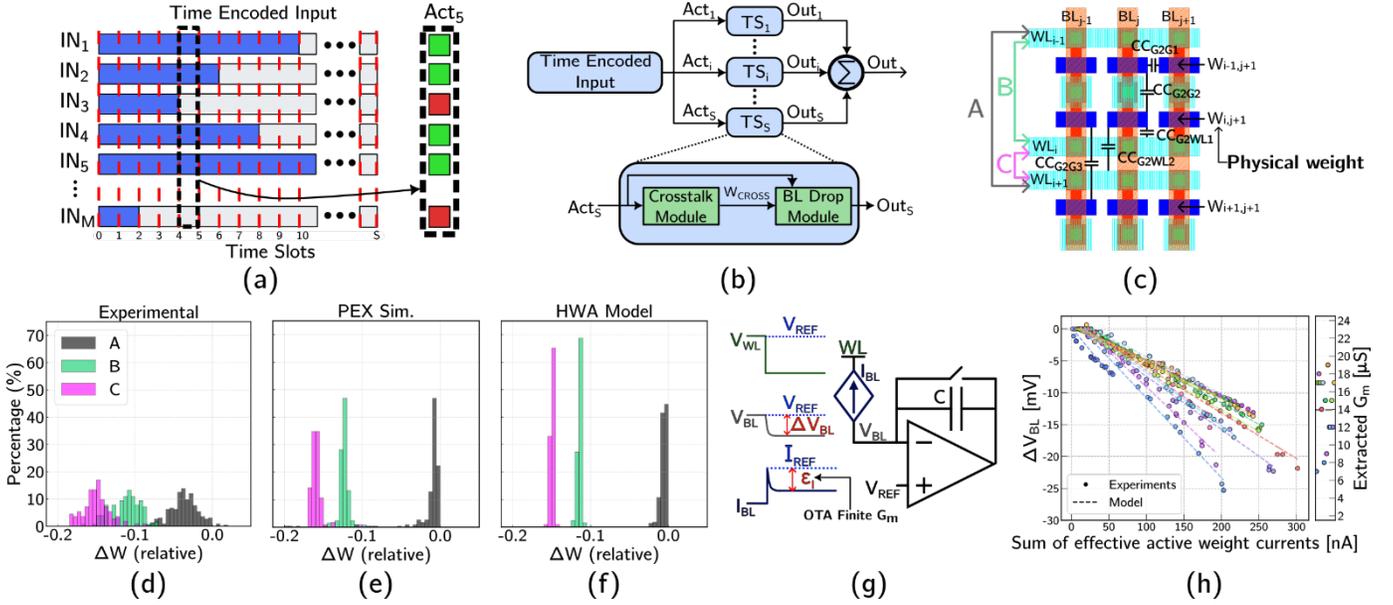

*Figure 2:* (a) Time diagram of an input pulse sample and related discretization within each time slot into the corresponding binary active vector (Act). (b) Hardware-aware (HWA) model block diagram. (c) Details of parasitic capacitances influencing the FG coupling factor and definition of crosstalk test scenarios: activation of two non-adjacent WLs (A), two far adjacent WLs (B), two close adjacent WLs (C). (d-f) Histogram of ΔW error calculated as the difference of the sum of two weights extracted with independent WL activation and the corresponding effective weight extracted with both WLs activated together: (d) experiments, (e) circuit simulations including parasitic capacitances, (f) HWA model simulations. (g) Circuit schematic illustrating the BL drop effect. (h) Experiment for the BL drop effect calibration.

### III. HARDWARE-AWARE VMM MODEL DESIGN AND CALIBRATION

*A. Proposed model architecture*

Due to the TD nature of the underlying accelerator, the effective distortion within each VMM operation depends on the temporal structure of the pulse-encoded inputs. To capture this behavior, the model discretizes each VMM operation into a set of time intervals, or time slots (TSs), as illustrated in Fig. 2a. The temporal discretization is determined through two complementary mechanisms. The model relies on a minimum time step simulation parameter, which sets the finest temporal resolution that can be represented. The maximum number of temporal simulation points is related to the minimum time step parameter, and it reflects the classical trade-off between accuracy and simulation time typical of transient circuit simulations: a smaller step increases temporal fidelity but lengthens simulation time, whereas a coarser step reduces computational effort while limiting the representation of time-dependent analog behavior. When the inputs are already quantized, the time resolution of the input data is automatically used as minimum time step of the transient model. This prevents unnecessary computations and ensures that only the relevant temporal intervals are represented. Due to this adaptive mechanism, the resulting TSs generally have non-uniform duration: each TS spans exactly the temporal region between two consecutive unique input levels.

For each TS, the model extracts a Boolean activation pattern (Act), indicating which inputs remain active within that interval (Fig. 2a). Crucially, the computation over all TSs does not require sequential simulation linked to the actual timing. Instead, the model processes all activation patterns using a fully vectorized formulation, enabling the GPU to evaluate all TS contributions in parallel. This reduces execution time substantially while leaving the total number of arithmetic operations unchanged.

As schematically shown in Fig. 2b, each activation pattern is processed through the two modules that model the dominant distortion mechanisms, i.e. **capacitive crosstalk** and **bit-line voltage drop**. Fig. 2b illustrates the overall flow of the TD-VMM model: the time-encoded input is decomposed into TS, each TS generates an activation vector, and each activation vector is fed through the distortion modules. The physical behavior and mathematical formulation of these modules are detailed in the following subsections. Finally, the contributions from all TSs are accumulated to obtain the effective output of the TD-VMM operation. The proposed HWA model is implemented with the *PyTorch* framework to allow seamless integration into modern deep learning workflows and efficient execution on GPU hardware.

*B. Crosstalk*

In a 1T-FG cell, the FG potential is activated through capacitive coupling, i.e., when a $V_{DS}$ voltage is supplied, a portion of this voltage results in a $V_{GS}$ increase, turning on the cell current. This portion of voltage depends on the coupling factor $k_{COUPL}$, which is the ratio of the gate-to-drain capacitance $C_{GD}$ to the total gate capacitance of the transistor $C_{GG}$, as illustrated in Fig. 1f. However, in a real crossbar array memory implementation, additional parasitic capacitances affect the total gate capacitance, inducing interactions between adjacent FGs and between each FG device and neighboring WLs, as shown in Fig. 2c.

To quantify the impact of the crosstalk, we have compared weights extracted under different conditions. First, given that the crosstalk effect depends on the activation of adjacent WLs, we take as a reference condition the case in which weights are extracted by activating each WL independently, one by one. Then, we have extracted the combined effect of two close weights belonging to the same BL through simultaneous activation of two WLs. We have considered three distinct cases, defined in Fig. 2c:

Case A: two WLs are activated while skipping one in between.
Case B: two consecutive WLs are activated, with the first one being an odd-indexed line.
Case C: two consecutive WLs are activated, with the first one being an even-indexed line.

Note that cases B and C differ due to the specific memory array layout: although the WLs in case B are consecutive in terms of indexing, they are physically farther apart, whereas in case C the two consecutive WLs are placed adjacent to each other in the layout. Under the idealized assumption of having a perfectly linear VMM, activating a pair of WLs should produce a weight equal to the sum of the two individual weights obtained by activating each WL independently. However, this condition is hardly met. To characterize the deviation in the experimental VMM, the ΔW difference was computed between the sum of the independently extracted weights and the "combined" weight obtained when the two WLs were activated together.

Figure 2d reports ΔW for the three scenarios A, B, and C. The largest ΔW is observed for case C, followed by B, and then A. This ordering is consistent with the physical spacing of the WLs in the layout: case C involves the closest WL pair, case B a slightly larger separation, and case A the largest spacing due to the skipped line. To extract an analytical model explaining this behavior we have exploited post-layout parasitic-extraction (PEX) simulations on the TD-VMM layout, using Calibre xRC. Beyond intrinsic device capacitances (i.e. $C_{GS}$, $C_{GD}$, $C_{GB}$), already included in the schematic-level transistor compact model, the total gate capacitance $C_{GG}$ must also include the parasitic capacitance between a transistor gate and nearby gates (e.g. $C_{G2G2}$ and $C_{G2G3}$ in Fig. 2c), as well as between a transistor gate and nearby WLs (e.g. $C_{G2WL1}$ and $C_{G2WL2}$ in Fig. 2c). Unfortunately, the parasitic extraction tool assumes that capacitances like $C_{G2WL2}$ are negligible, due to the presence of an intermediate metal conductor that is (wrongly) assumed to totally shadow the gate from the other metal layers. They are not: therefore we have performed an estimate of $C_{G2WL2}$ by relying on a finite-element simulation performed with COMSOL Multiphysics.

Considering all extracted capacitances, in Fig. 3e we have repeated by simulation the same analysis as in Fig. 3d (which instead was based on experimental data), confirming that the observed behavior is indeed layout-related and can be attributed to crosstalk effects. Although the resulting distributions are more compact than the experimental ones (device mismatch is not simulated), the overall trend is fully consistent with experiments, with ΔWs extracted in case A negligible compared to those in cases B and C. By carrying out simple analytical calculations, based on an exponential subthreshold transistor model and the relevant capacitive coupling effects, it can be shown that the weight reduction induced by crosstalk, for each layout scenario, results in a purely linear reduction, according to:

$$W_{with\ XT} = \alpha \cdot W_{wo.\ XT}, \qquad (2)$$

where the coefficient $\alpha < 1$ can be expressed as:

$$\alpha = e^{\frac{V_{DS} \cdot (k_0 - k_{XT})}{\eta \cdot V_T}} \qquad (3)$$

where $V_T$ is the thermal voltage, $\eta$ is the subthreshold slope factor, and $k_0$ and $k_{XT}$ are the $V_{DS}$ to FG coupling factors without and with the crosstalk effect, respectively. This is attributed to the fact that, for any $W_{i,j}$ weight-storing cell, the activation of an adjacent WL (e.g. WL$_{i+1}$) changes the coupling factor between the WL$_i$ and the respective transistor $M_{i,j}$. We have modeled this effect only for adjacent WLs. By simulating all the combinations of two adjacent WLs activation, we have found only two α distinct values, corresponding to the cases B and C, respectively, confirming that the weight reduction is linear and layout related.

It should be noted that the analysis in Fig. 2d and 2e mainly reflects the impact of crosstalk, while the BL-drop contribution will be discussed in the following subsection. Since crosstalk is the dominant mechanism for the behavior examined here, Fig. 2f reports the results of a similar analysis conducted in Fig. 2d and 2e, obtained using the full version of the extracted HWA model, which includes both crosstalk and BL-drop effects. Even if the complete model has not yet been introduced, anticipating this result is useful, as it confirms that the model faithfully reproduces the crosstalk-related trends observed experimentally and in the PEX simulations.

*C. BL drop*

As already introduced, the current of a 1T–FG cell depends on the $V_{DS}$, which in turn defines the $V_{GS}$ through capacitive coupling. In the proposed VMM, the WL voltage serves as the source potential and it is actively driven to 0 V by the dedicated WL driver enabling row selection. The drain node corresponds to the BL, whose potential is stabilized at $V_{REF}$ by an operational transconductance amplifier (OTA). The OTA must not only maintain this reference voltage but also provide the total BL current, $I_{BL}$. The relation between the current and the differential input voltage can be expressed as:

$$I_{BL} = G_m v_{id} \qquad (4)$$

where $G_m$ is the OTA transconductance and $v_{id}$ its differential input voltage. This implies that, to sustain a certain current, the BL voltage must drop. Such a reduction lowers the effective $V_{DS}$ of the cells and, consequently, their active current. A schematic illustration of this limitation is shown in Fig. 3g. When one or more WLs are activated, the expected BL current should ideally rise to a reference value $I_{REF}$, theoretically defined as:

$$I_{REF} = \sum_i^{Active\ WLs} I_0 \exp\left(\frac{k_{COUPL} \cdot V_{REF} - V_{th,i}}{\eta \cdot V_T}\right), \quad (5)$$

where $I_0$ is the specific current and $V_{th,i}$ is the threshold voltage, also accounting for the shift induced by the stored charge defining the weight.

This leads to a deviation between the theoretical reference current and the actual BL current, which we denote in Fig. 2g as the error $\varepsilon_I$. Determining the actual BL voltage requires solving the following transcendental equation:

$$-G_m \cdot \Delta V_{BL} = I_{REF} \cdot \exp\left(\frac{k_{COUPL} \cdot \Delta V_{BL}}{\eta \cdot V_T}\right). \quad (6)$$

To improve computational efficiency, this solution is replaced by a look-up table (LUT), which directly estimates the BL voltage drop from the reference current, which corresponds to the sum of the crosstalk-aware active weight currents (provided by the dedicated module). The corresponding relative error during TS integration is evaluated as:

$$\varepsilon_r = \exp\left(\frac{k_{COUPL} \cdot \Delta V_{BL}}{\eta \cdot V_T}\right) - 1. \quad (7)$$

To experimentally validate and calibrate this effect, we performed the following measurement procedure: by keeping the reset transistor active, the Miller integrator was configured as a buffer, allowing direct measurement of the BL voltage at its output. Subsequently, different WLs were activated in sequence, and the corresponding BL voltages were recorded as a function of the effective sum of the active weights, as reported in Fig. 2h. These measurements reveal that different BLs exhibit distinct voltage–current characteristics, even for identical nominal weight currents.

The slopes of the best-fit lines, which correspond to the transconductance $G_m$, vary among columns (represented by different colors), primarily due to process-induced mismatches in the fabricated OTAs. Since these variations are not included in the present model, a fixed transconductance value of $G_m$ of $14\ \mu S$, corresponding to the median of the extracted values, was adopted as a reference value in the following analysis.

## IV. WEIGHT EXTRACTION

In typical analog computing workflows, network training is performed offline using software models executed on high-performance computing servers. The resulting weight matrix is then transferred to the analog hardware for low-power inference. This paradigm places a strong emphasis on accurate weight programming, as inaccurate weights can degrade classification accuracy. Since the programming process relies on a program-and-verify scheme [11], the ability to correctly extract the weights becomes fundamental: a weight can accurately be written only if it can be accurately read. Thus, the extraction methodology can be a bottleneck for the overall system performance.

However, due to analog non-idealities, such as crosstalk, finite $G_m$, leakage, and circuit mismatch, different weight extraction techniques can provide different values for the weight associated to the same 1T-FG cell. As a result, the notion of a "true" (or "effective") weight becomes context-dependent, and a careful methodology is required to ensure consistent and accurate measurements. To explore this, we analyzed four different Weight Extraction Methods (WEM) based on four different input datasets:

1. **WEM1, single word-line**: each input $WL_i$ is activated individually while all other WLs are held inactive. A series of experiments is performed in which the $i$-th input pulse width is incrementally varied across measurements, following a ramp-like progression. The values of $\Delta V_{out_j}$ are recorded for each input condition as shown in Fig. 3a.

2. **WEM2, all word-lines but one**: all word-lines except one are simultaneously activated with the same pulse, while a single line is kept inactive. Fig. 3b illustrates how the input pulse widths are swept in a ramp-like progression across experiments, leading to ramp-shaped outputs. The linear region of the output is then analyzed to estimate the contribution of the inactive line by subtraction.

3. **WEM3, random inputs**: 16-elements input vectors are sampled from a uniform random distribution. Fig. 3c shows the output distribution for a batch of 512 inputs.

4. **WEM4, low-resolution MNIST image stimuli**: input vectors are derived from $4 \times 4$ pixel images downsampled from the MNIST dataset. As in the random case, the TD-VMM outputs are collected, and the distribution is shown in Fig. 3d.

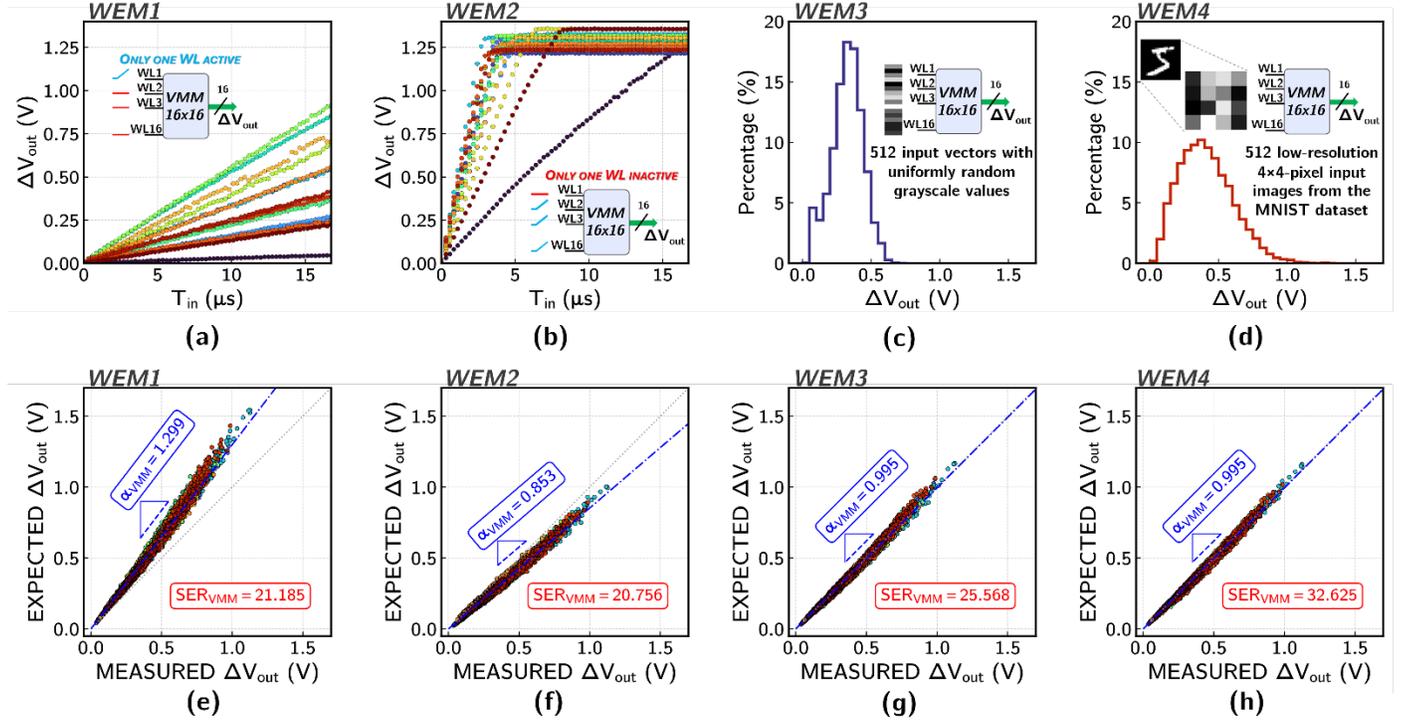

*Figure 3:* Weight extraction methods (WEMs). (a) WEM1, single word line: a set of ramp-like input pulse widths is applied to one WL at a time, while all other WLs are held inactive. (b) WEM2, all WLs but one: ramp-like inputs are applied simultaneously to all WLs except one, which remains inactive. (c) WEM3, random input images: 512 vectors of 16 elements sampled from a uniform random distribution. (d) WEM4, low-resolution MNIST images: 512 input vectors derived from 4×4 pixel images sampled from the MNIST dataset. (e-h) TD-VMM tested with low-resolution MNIST images (similar to WEM4 experiment, but with different images): scatter plots of theoretically expected bitline output voltage, calculated using the real inputs and the weights extracted with the corresponding WEM experiment, versus actually measured bitline output voltage. Reported parameters: slope of the linear fit ($\alpha_{VMM}$), signal-to-error ratio ($SER_{VMM}$).

In all cases, the corresponding output voltages are used to estimate the weights through regression based on two different input-output models: (i) the ideal VMM model and (ii) the HWA model, accounting for capacitive crosstalk and BL voltage drop. For WEM1 and WEM2, only output values below the saturation region of the OTA are considered during weight extraction, whereas for WEM3 and WEM4 the input data are rescaled so that 99.7% of the resulting $\Delta V_{out,j}$ values remain below the saturation level. After weight extraction, we have tested again the same VMM using a new set of inputs and collected the resulting output voltages. Based on the 4 sets of weights, extracted under the various WEMs, we have also computed the theoretically expected outputs for each case, as the product of the actual inputs by the extracted weights. By plotting the scatter plots of theoretically expected outputs as a function of actually measured outputs, we have extracted two parameters, which are the best linear fitting slope ($\alpha_{VMM}$) and the Signal-to-Error Ratio (SER), which is derived by considering the ratio of the rms theoretical outputs to the rms of the error (calculated after linear fit correction).

*A. Weight extraction with ideal VMM model*

The analysis of the scatter plots in Fig. 3e–h reveals how different extraction methodologies influence the consistency of the estimated weights in the TD-VMM array. In all cases, the $x$-axis corresponds to the same set of measured TD-VMM outputs, obtained by applying a fixed test batch of 512 MNIST images to the chip. This ensures a fair and consistent comparison across all extraction methods. Note also that, in WEM4, the MNIST batch used for weight extraction is different from the one used for testing. WEM1 (*single word-line ramp*) provides a peculiar condition for weight extraction: since only one WL is activated at a time, while all others remain inactive, the BL currents reflects the contribution of single cells. This condition minimizes the total current flowing through the BL and thus represents a favorable scenario from the BL voltage drop viewpoint. Furthermore, since no adjacent WLs are simultaneously active, capacitive crosstalk effects are inherently suppressed. WEM2 (all word-lines active except one) represents the opposite extreme: almost all WLs are simultaneously pulsed, resulting in strong capacitive coupling between adjacent cells and unfavorable scenario for BL current and corresponding BL voltage drop. During inference with random-pixels images (WEM3) and low-resolution MNIST stimuli (WEM4), the operating point lies between these two conditions. Only a subset of WLs is typically active, producing intermediate levels of crosstalk and BL drop. Among the investigated approaches, WEM3 and WEM4 yield the best quantitative results, with slope values of $\alpha_{VMM} = 0.995$ for both, very close to the ideal value of 1, and SER of 25.6 and 32.6, respectively. In particular, the WEM4 MNIST-based method most closely reproduces the actual input statistics encountered during inference, providing the most representative operating conditions for the VMM circuit.

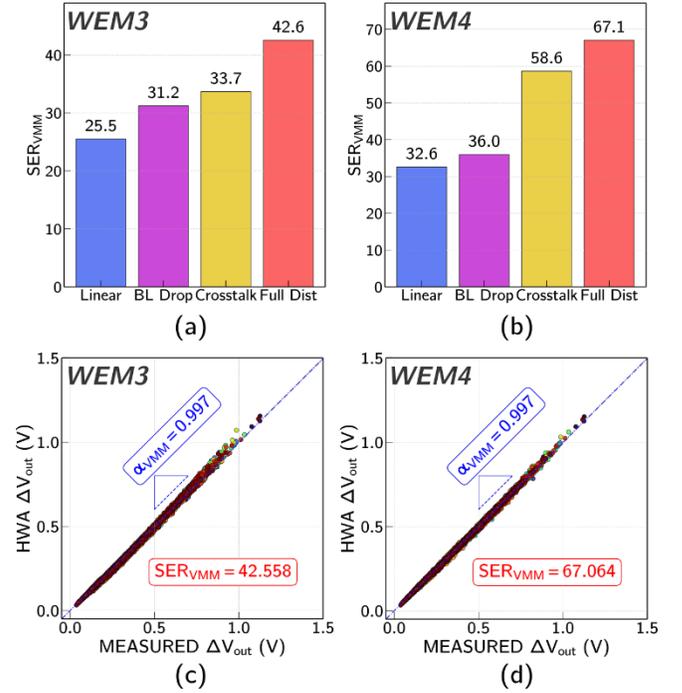

*Figure 4:* $SER_{VMM}$ improvement for weights extracted by stimulating the VMM according to WEM3 (a) and WEM4 (b) strategies. Four models are considered to extract he expected outputs: ideal VMM, HWA model including only the BL-drop effect, HWA model including only the crosstalk effect, full-HWA model. Scatter plots considering the full-HWA model for WEM3 (c) and WEM4 (d).

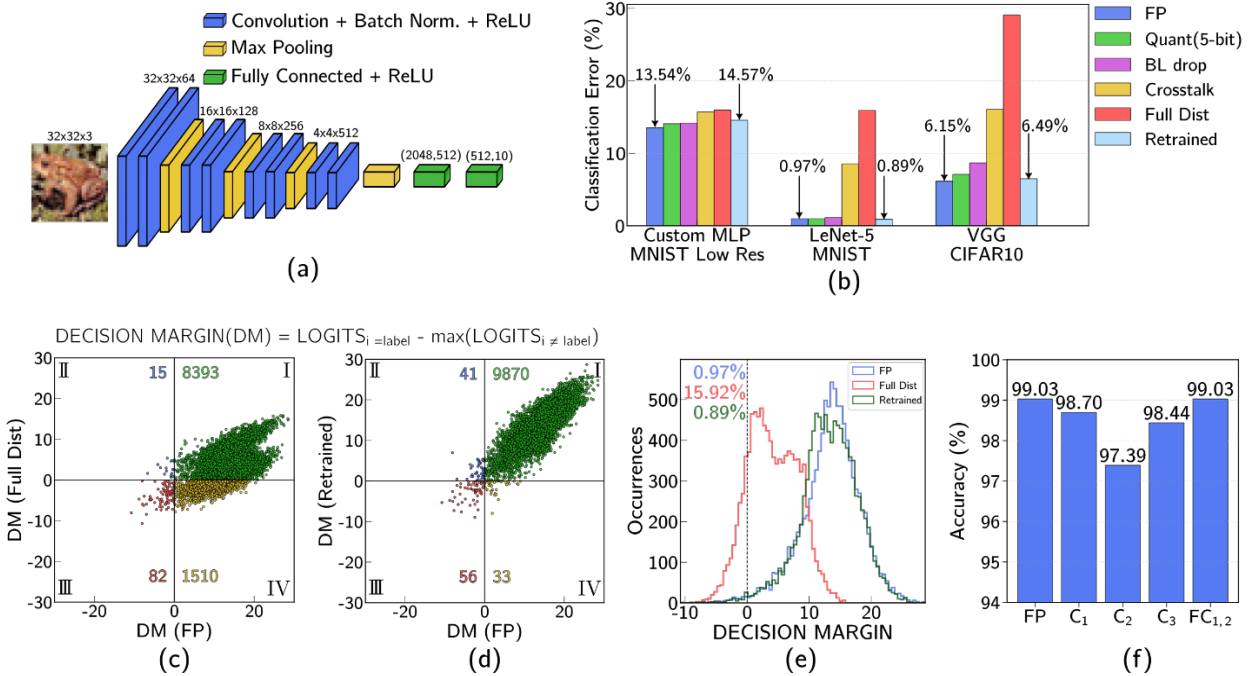

*Figure 5:* (a) Architecture of the VGG-style convolutional neural network used for CIFAR-10 classification. The model consists of a sequence of convolutional blocks with batch normalization and ReLU activation, followed by two fully connected layers. (b) Custom MLP, LeNet-5, and VGG-style, inference accuracy across the full evaluation pipeline: ideal neural network model with floating point arithmethics, 5-bit I/O quantization, impact of BL voltage drop evaluated with HWA model, impact of capacitive crosstalk evaluated with HWA model, impact of both non-idealities computed with full HWA model, and finally impact of full-HWA model retraining. (c) LeNet-5 scatter plot of Decision Margin (DM) obtained with the HWA model and the ideal network model with floating point arithmethic (DM: difference between the logit of the correct label and the maximum logit of the other classes. Each correct inference leads to a positive DM). (d) Improvement of the HWA model DM after retraining. (e) Histograms of the DM for the LeNet-5 model (same data used for c and d). (f) Impact of the hardware nonidealities on the LeNet-5 inference accuracy when the hardware VMM is used to implement individual layers.

Consequently, this approach achieves the highest SER, confirming that extracting the weights using the same type of dataset employed during inference allows the results to reflect realistic hardware conditions and ensures the most accurate estimation.

*B. Weight extraction with hardware-aware model*

In the previous subsection, four extraction methods were compared using the ideal VMM model, and it was observed that WEM3 and WEM4 methods yield the best results in terms of SER. The same experimental data were processed to extract the weights by using the proposed HWA model. For each WEM dataset, we have extracted the weights using two reduced version of the HWA model, where either only the BL drop or only the crosstalk module was considered, as well as with the full HWA version. Fig. 4a and 4b, which refer to WEM3 and WEM4 weight families, show the improvement of the SER moving from weights extracted with the ideal model (25.5 and 32.6), to weights extracted with BL-drop-only HWA model (31.2 and 36) and weights extracted with crosstalk-only HWA model (33.7 and 58.6), reaching the maximum when the weights are extracted with the full HWA model (42.6 and 67.1). The scatter plot related to full HWA WEM3 and WEM4 extractions are reported in Figs. 4c and d, respectively (note the reduced dispersions with respect to the corresponding ones extracted with theoretical model in Figs. 3g and 3h, respectively). It is important to emphasize that, although crosstalk appears to play a more fundamental role than BL drop, this should not be interpreted as an absolute conclusion. In fact, the observed behavior is strongly dependent on: (1) the test dataset used, as it could be a dataset that might stress capacitive coupling between adjacent WLs more than the BL-drop, due to maximum BL current remaining low; and (2) the fact that the relative influence of these factors may shift as the VMM size scales. Noticeably, combining both modules in the full model further increased the SER of WEM4 case to 67.1, effectively doubling the SER achieved with the best theoretical VMM model extraction shown in Fig. 3(h).

Overall, the MNIST-based extraction remains the most effective approach, highlighting the importance of aligning the weight-extraction procedure with the operating conditions expected during inference. Using the same type of dataset employed in actual inference, such as low resolution MNIST, ensures that the extracted weights reflect realistic input statistics and device interactions, thereby maximizing consistency between software-trained parameters and hardware-implemented behavior. Moreover, adopting the HWA for regression significantly enhances extraction accuracy: by explicitly accounting for capacitive crosstalk and BL voltage drop, the resulting SER is approximately doubled compared to the theoretical linear model. This improvement enables a more precise estimation of the effective weights, providing a more reliable reference for subsequent program-and-verify operations and ultimately improving the fidelity of hardware weight programming.

## V. Hardware-aware Training

To evaluate the impact of the proposed HWA TD-VMM model on neural network performance, we considered three architectures of increasing complexity: (i) a custom two-layer MLP structurally matched to the fabricated TD-VMM and operating on 4×4 MNIST inputs (Fig.1c); (ii) the LeNet-5 convolutional network trained on full-resolution MNIST; and (iii) a VGG-style convolutional network trained on CIFAR-10, illustrated in Fig. 5a. This progression spans from very small input dimensionality to deeper, data-intensive models, enabling a comprehensive assessment of distortion sensitivity. For each architecture, the evaluation followed a common procedure. We began from the ideal neural network with floating-point (FP) arithmetics. When then applied 5-bit quantization at the input and output of each layer to emulate the effect of on-chip digital-to-analog and analog-to-digital converters (5-bit I/O quant.). We then evaluate the impact on accuracy of the bit-line voltage drop, of the capacitive cross talk, and of both effects concurrently, using the hardware-aware TD-VMM model. The resulting inference accuracies are summarized in Fig. 5b. Crosstalk emerges as the dominant source of degradation across all models, while bit-line voltage drop alone has a limited effect. Importantly, the accuracy drop observed when both non-idealities are applied simultaneously is larger than the sum of the individual degradations, indicating that the two mechanisms interact in a non-linear fashion rather than contributing independently. This combined distortion becomes increasingly severe as the network complexity grows: although the custom MLP experiences only a moderate increase of classification error (+1.9%), both the LeNet-5 on standard MNIST (+14.95%) and the VGG-style network on CIFAR-10 (+22%) undergo a substantial error increase.

We show the effectiveness of HWA retraining, performing an epoch training using the HWA model in the forward propagation and the ideal model in the backpropagation, using the weights computed in the first training of the ideal model. With this approximation, the backward pass avoids the complexity of differentiating cost function of the HWA model, and reaps the full benefits of its forward-path accuracy.

As shown in Fig. 5b, hardware-aware retraining restores performance across all three architectures. The custom MLP almost entirely recovers its 5-bit quantized baseline. LeNet-5 and the VGG-style network not only recover the distortion-induced accuracy loss, but even surpass the accuracy of their quantized baselines. In these two cases, the networks clearly benefit from being simultaneously aware of both analog distortions and quantization constraints, learning parameters that are inherently more resilient than those obtained through standard floating-point training followed by quantization. These results demonstrate that the proposed methodology scales effectively from small to deep networks and that HWA training can substantially improve robustness.

To further clarify the origin of the accuracy degradation, we performed a more detailed investigation on the LeNet-5 network, examining how TD-VMM distortions affect individual activations and decision confidence. Fig. 5c and 5d report a set of complementary visualizations based on the "*Decision Margin*" (DM), here defined as the difference between the logit associated with the correct class and the highest competing logit. The scatter plot in Fig. 5c compares the DMs produced by the HWA model before retraining and the ideal neural network. A large concentration of points appears in the fourth quadrant, indicating samples correctly classified in floating point but misclassified once TD-VMM distortions are applied. This quadrant provides a direct visualization of the classification inconsistencies introduced by BL voltage drop and crosstalk.

Retraining effectively resolves these inconsistencies, as shown in Fig. 5d: after retraining the HWA model has much improved DMs: the scatter plot becomes much more tightly concentrated around the diagonal, and the population of points in the fourth quadrant is greatly reduced compared to the pre-retraining case. Moreover, the remaining points are now comparable in number to those in the second quadrant, where the retrained network correctly classifies samples that the analog network before retraining misclassifies. Overall, since the points in the second and fourth quadrants are more than two orders of magnitude fewer than those in the first, this demonstrates that HWA training restores agreement between the distorted and ideal models on a sample-by-sample basis, not just at the aggregate accuracy level. The difference of the DM distributions is further illustrated in Fig. 5e. Hardware distortions significantly compress and shift the DM distribution toward zero and beyond toward negative values, consistently with the observed accuracy loss. After retraining, the distribution realigns with the floating-point case, confirming that the network has learned parameters that are robust to both crosstalk, BL voltage drop and 5-bit quantization of input data.

Finally, Fig. 5f shows the impact of selectively considering the non-idealities of individual layers of the LeNeT-5, using the the HWA model. When applied independently to each of the first three convolutional layers, an appreciable accuracy drop emerges, which is substantially larger than the degradation observed when only the final fully connected layer is distorted. Nevertheless, the results do not indicate that any single layer accounts for the full loss observed when distortions are applied globally. Instead, the degradation arises from a combined effect across the network, where individual layer-level losses accumulate through successive stages.

## VI. Conclusion

This work establishes a physics-informed hardware-aware training framework tailored for time-domain analog in-memory computing, bridging the gap between idealized software models and real-world hardware behavior. By accurately modelling crosstalk and bit-line voltage drop, and validating against silicon measurements, we demonstrate that retraining with the proposed model restores baseline accuracy across diverse neural architectures. These results confirm that time-domain analog accelerators can achieve competitive performance without sacrificing energy efficiency, paving the way for scalable, deployment-ready AIMC solutions in edge and embedded AI systems.

**Giulio Filippeschi** received the MS degree in Electrical Enegineering from the University of Pisa in 2024. He is now pursuing is Ph.D. at the University of Pisa with a research activity on neuromorphic systems design.

**Mirko Brazzini** received the MS degree in Electrical Enegineering from the University of Pisa in 2024. He is now pursuing is Ph.D. at the University of Pisa with a research activity on neuromorphic systems design.

**Cristhopher Mosquera** received the MS degree in Electrical Enegineering from the University of Calabria in 2024. He is now pursuing is Ph.D. at the University of Calabria in collaboration with Quantavis s.r.l. with a research activity on neuromorphic systems design.

**Marco Lanuzza** (Senior Member, IEEE) received the Ph.D. degree in electronic engineering from the Mediterranea University of Reggio Calabria, Reggio Calabria, Italy, in 2005.,Since 2006, he has been with the University of Calabria, Rende, Italy, where he is currently an Associate Professor. He has authored over 150 publications in international journals and conference proceedings. His recent research interests include the design of ultralow voltage circuits and systems, the development of efficient models and methodologies for leakage- and variability-aware designs, and the design of digital and analog circuits in emerging technologies.,Prof. Lanuzza is an Associate Editor of Integration, the VLSI Journal.

**Alessandro Catania** (Senior Member, IEEE) received the B.S., M.S., and Ph.D. degrees in electronic engineering from the University of Pisa, Italy, in 2014, 2016 and 2020, respectively. He is currently working as an Assistant Professor with the University of Pisa. His current research interests include mixed-signal microelectronic design for harsh environments and wireless power transfer systems for implantable systems.

**Sebastiano Strangio** (Senior Member, IEEE) received the B.S. and M.S. degrees (cum laude) in EE, and the Ph.D. degree from the University of Calabria, Cosenza, Italy, in 2010, 2012, and 2016, respectively. He was with IMEC, Leuven, Belgium, as a Visiting Student, in 2012, working on the electrical characterization of resistive-RAM memory cells. From 2013 to 2016, he was a Temporary Research Associate with the University of Udine, and with the Forschungszentrum Jülich, Germany, as a Visiting Researcher, in 2015, researching on TCAD simulations, design, and characterization of TFET-based circuits. From 2016 to 2019, he was with LFoundry, Avezzano, Italy, where he worked as a Research and Development Process Integration and Device/TCAD Engineer, with main focus on the development of a CMOS Image Sensor Technology Platform. He is currently Associate Professor of Electronics with the University of Pisa. He has authored or coauthored over 40 papers, most of them published in IEEE journals and conference proceedings. His research interests include technologies for innovative devices (e.g. TFETs) and circuits for innovative applications (CMOS analog building blocks for DNNs), as well as CMOS image sensors, power devices and circuits based on wide-bandgap materials.

**Giuseppe Iannaccone** (Fellow, IEEE) received the M.S. and Ph.D. degrees in EE from the University of Pisa, in 1992 and 1996, respectively. He is currently Deputy President and Professor of electronics with the University of Pisa. He has coordinated several European and national projects involving multiple partners and has acted as principal investigator in several research projects funded by public agencies at the European and national level, and by private organizations. He co-founded the academic spinoff Quantavis s.r.l. and is involved in other technology transfer initiatives. He has authored or coauthored more than 250 articles published in peer-reviewed journals and more than 160 papers in proceedings of international conferences, gathering more than 13,000 citations on the Scopus database. His research interests include quantum transport and noise in nanoelectronic and mesoscopic devices, development of device modeling tools, new device concepts and circuits beyond CMOS technology for artificial intelligence, cybersecurity, implantable biomedical sensors, and the Internet of Things. He is a fellow of the American Physical Society.